\documentclass{aastex}
\usepackage{spr-astr-addons}

\RequirePackage{color}

\begin{document}

\title{Investigation of the Relationship Between the Air Pollution and Solar Activity}
\slugcomment{Not to appear in Nonlearned J., 45.}
\shorttitle{Air Quality and Solar Activity} \shortauthors{Tan et
al.}

\author{Chengming Tan\altaffilmark{1,2,3}, Baolin Tan\altaffilmark{1,2,3}, Bisong Liu\altaffilmark{4}}

\affil{$^{1}$CAS Key Laboratory of Solar Activity, National
Astronomical Observatories of Chinese Academy of Sciences, Datun
Road A20, Chaoyang District, Beijing 100012, China}

\affil{$^{2}$Sate Key Laboratory of Space Weather, Chinese Academy
of Sciences, Beijing 100190, China}

\affil{$^{3}$School of Astronomy and Space Sciences, University of
Chinese Academy of Sciences, Beijing 100049, China}

\affil{$^{4}$Beidou Intelligence Information and Network
Technology Limited Company, Beijing 100041, China}


\begin{abstract} How did the Sun affect the air
pollution on the Earth? There are few papers about this question.
This work investigates the relationship between the air pollution
and solar activity by using the geophysical and environmental data
during the period of 2000-2016. It is quite certain that the solar
activity may impact on the air pollution, but the relationship is
very weak and indirect. The Pearson correlation, Spearman rank
correlation, Kendalls rank correlation, and conditional
probability were adopted to analyze the air pollution index (API),
air quality index (AQI), sunspot number (SSN), radio flux at
wavelength of 10.7 cm (F10.7), and total solar irradiance (TSI).
The analysis implies that the correlation coefficient between API
and SSN is weak ($-0.17<r<0.32$) with complex variation. The main
results are: (1) For cities with higher air pollution, the
probability of high API will be increased along with SSN, then
reach to a maximum, and then decreased; (2) For cities with lower
air pollution, the API has lower correlation with SSN; (3) The
relationship between API and F10.7, or API and TSI are also
similar as API and SSN. The solar activities take direct effect on
TSI and the energetic particle flux, and indirect and long-term
effect on lower atmosphere and weather near the Earth. All of
these factors contribute to the air pollution on the Earth.
\end{abstract}


\section{Introduction}

The air pollution impact on human health greatly in many ways,
such as seeing, breathing, coughing, pulmonary disease, cancer,
and so on \citep{Sicard11, Cooper12}. Many people believed that
the air pollution should be produced mainly from industry,
automobiles and domestic fuels fire, cooking, smoking and so on.
\citep{Cooper12} provided the worldwide distribution of pollution
from local and global sources. The heavy polluted area are
spreaded all over the industrial estate and populous zone.

Then, are there some external contributors to the cities air
pollution? Because the solar irradiance dominates the total energy
acting on the Earth atmosphere, and solar activities vary the
total solar irradiance (TSI). Therefore, it is reasonable to
suppose that the solar activities may affect on the Earth climate,
air ventilation, and even the air pollution \citep{Pudovkin04,
Kutiev13}. Actually, the solar activities affect ionization in the
Earth atmosphere, as well as the lower atmosphere state, weather
conditions, the distribution of ozone molecules and urban
environment, and therefore the air pollution on the earth.
However, so far, the relationship between solar activity and air
pollution is still not quite certain and have been studied rarely
\citep{Sharma97, Mavrakis08}. \citep{Sharma97} only deduced and
discussed the influence of air pollution theoretically from the
Sun, cosmic ray ionization, and lightening activity.
\citep{Mavrakis08} studied the air quality of Thriassion Plain --
Greece covering more than 23 years. But he did not give a certain
results or conclusion on the relationship between solar activity
and air quality.

This work investigated the relationship between solar activities
and the records of air pollution in several typical Chinese big
cities. Does the solar activities impact on the cities air
pollution? How much is the level of the impact? What is the cause
of the relationship? In order to answer these question, we analyze
different correlation and mainly conditional probability among the
air pollution index and several solar activity index, such as the
sunspot number (SSN), solar radio flux at wavelength of 10.7 cm
(F10.7), and the total solar irradiance (TSI). In Section 2, we
analyze the record data in three typical Chinese big cities, and
then analyze globally the data of many cities of three classes.
The conclusion and discussion are summarized in Section 3.

\section{Data and analysis}

\subsection{Data Source}

The data involving in this research includes the air pollution
index (API) of tens of Chinese cities and several solar activity
indexes. The later includes sunspot number (SSN), solar radio flux
at wavelength of 10.7 cm (F10.7), and the total solar irradiance
(TSI). This work focused on investigating the relationship between
API and the solar activity indexes.

API data is recorded by the Chinese Ministry of Environmental
Protection (MEP) from 2000 to 2013. The API level is based on the
level of 6 atmospheric pollutants, namely sulfur dioxide
(SO$_{2}$), nitrogen dioxide (NO$_{2}$), suspended particulates
smaller than 10$\mu$m and 2.5 $\mu$m in aerodynamic diameter
(PM10, PM2.5), carbon monoxide (CO), and ozone (O$_{3}$) measured
at the monitoring stations throughout all over the country.
Please note that only a few cities included the index of PM2.5, O$_{3}$, and CO.
After 1st January 2014, MEP monitors daily air quality level in hundreds
of its major cities and adopt Air Quality Index (AQI). AQI is an
index used by government agencies to communicate to the public how
polluted the air currently is or how polluted it is forecast to
become. An individual score of AQI (IAQI) is assigned to the level
of each pollutant and the final AQI is the highest of those 6
scores. The data of API and AQI are obtained from the data center
website\footnote{http://datacenter.mep.gov.cn/index} of the
Chinese MEP. The weather data are download from the data center of
China meteorological administration\footnote{http://data.cma.cn/}
and the weather website\footnote{http://lishi.tianqi.com/}.

The data of SSN and F10.7 are obtained from the
website\footnote{https://www.ngdc.noaa.gov/} of National Centers
for Environmental Information (NCEI). The data of TSI at 1AU is
from
ACRIMSAT/ACRIM3\footnote{ftp://ftp.ngdc.noaa.gov/STP/SOLAR\underline{\hspace{0.5em}}DATA/}
\citep{Willson14} and
SORCE/TIM\footnote{http://lasp.colorado.edu/home/sorce/data/}
\citep{Kopp05a, Kopp05b}. The Active Cavity Radiometer Irradiance
Monitor Satellite (ACRIMSAT) is a defunct satellite and instrument
that was one of the 21 observational components of NASA's Earth
Observing System program. The ACRIM3 instrument monitored total
solar irradiance (TSI). The Solar Radiation and Climate Experiment
(SORCE) is a NASA-sponsored satellite mission that is providing
state-of-the-art measurements of incoming x-ray, ultraviolet,
visible, near-infrared, and total solar radiation. Total
Irradiance Monitor (TIM) is one of the four instruments.

\begin{figure*}[ht]
\epsscale{1.50} \plotone{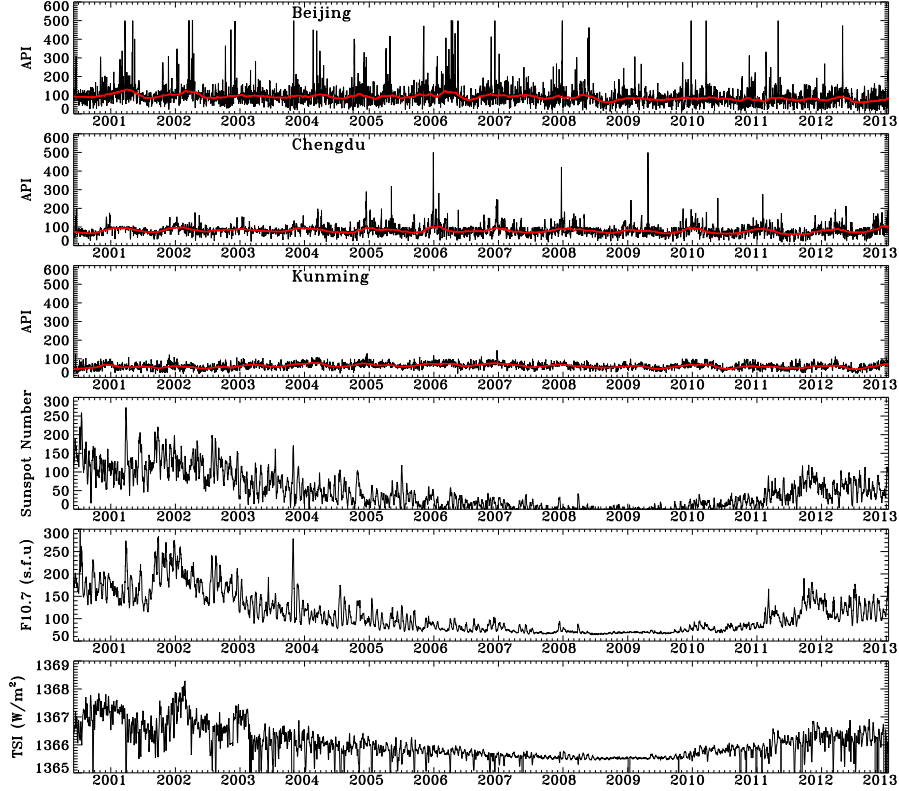} \caption{Upper three panels
plotted API in Beijing, Chengdu, and Kunming during 2000-2013,
respectively. The red line is the 101 points moving average after
excluding the large API. Bottom three panels plotted SSN, F10.7,
and TSI during 2000-2013, respectively. \label{figapi}}
\end{figure*}

\begin{table*}
\footnotesize
 \caption{PCC, SCC, and KCC of various
correlation of three cities as example\label{tbl-1}}
\begin{tabular}{@{}crrrrrrrrr@{}}
\tableline
City & API$\&$SSN & API$\&$SSN & API$\&$SSN & API$\&$F10.7 & API$\&$F10.7 & API$\&$F10.7 & API$\&$TSI & API$\&$TSI & API$\&$TSI  \\
\tableline
     & PCC & SCC & KCC & PCC & SCC & KCC & PCC & SCC & KCC  \\
\tableline Beijing & 0.084 & 0.094 & 0.061 & 0.10  & 0.11 & 0.074 & 0.079 & 0.097 & 0.064 \\
\tableline
Chengdu & -0.002 & 0.061 & 0.042 & 0.026 & 0.097 & 0.065 & 0.031 & 0.071 & 0.048 \\
\tableline
Kunming & -0.027 & -0.009 & -0.0057 & -0.07 & -0.047 & -0.031 & -0.13 & -0.085 & -0.056\\
\tableline
\end{tabular}
\end{table*}

\begin{figure*}[ht]
\epsscale{1.50} \plotone{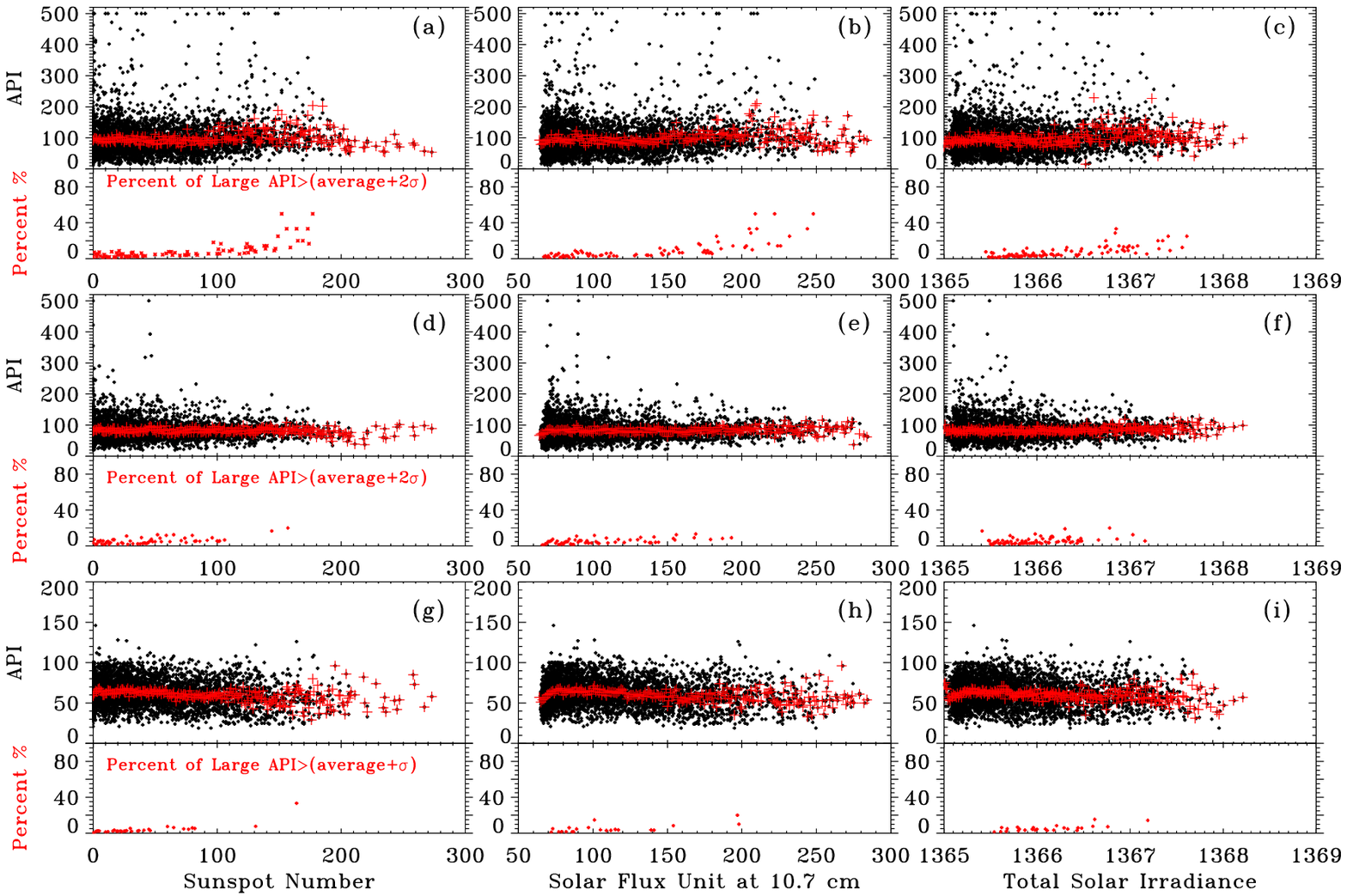}
\caption{The figure plotted for three cities the scatter diagram of
API versus SSN, API versus F10.7, API versus TSI, and percent of
large API, respectively. Upper part of panel
(a) plotted the scatter diagram of API versus SSN. The red cross
is the API average value at given SSN. Bottom panel of (a) plotted
the percent of large API ($>$average+2$\sigma$).\label{figsca}}
\end{figure*}

\subsection{The profile and characteristic of API in single city}

At first, we analyze the relationship between API and SSN, F10.7,
and TSI, respectively in 3 classes. The 3 classes are classified
according to the API distribution in the city. The first class is
the city with $\geq0.55\%$ days (more than two days per year
averagely) of large API ($\geq200$), for example, Beijing. The
second class is the city with $0.1\%\leq API<0.55\%$ days of large
API ($\geq200$), for example, Chengdu. The third class is the city
with $<0.1\%$ days of large API ($\geq200$), for example, Kunming.
The correlation of API and SSN, API and F10.7, API and TSI are
analyzed. In order to make a comparison, we adopted Pearson
correlation, Spearman rank correlation, and Kendalls rank
correlation. Pearson correlation coefficient (PCC) is a measure of
the linear correlation between two variables. Spearman rank
correlation coefficient (SCC) is a nonparametric measure of rank
correlation between two variables. Kendall rank correlation
coefficient (KCC) is statistically used to measure the ordinal
association between two measured quantities.

Upper three panels of fig.\ref{figapi} plotted the API in Beijing,
Chengdu, and Kunming of 4607 days during 2000-2013, respectively.
The red line went through the API curve is the 101 points moving
average after excluding the large API. We firstly exclude the
point with 2 times of standard deviation larger than the average.
Then, each point of the red line is the average value of $\leq101$
points nearby, ie, $\leq50$ points before and after. The large API
($\geq200$) in Beijing, Chengdu, and Kunming are about $3.19\%$
days, $0.48\%$ days, and no day, respectively. The bottom three
panels of fig.\ref{figapi} plotted SSN, F10.7, and TSI during
2000-2013, respectively. Table.\ref{tbl-1} is the PCC, SCC, and
KCC of API and SSN, API and F10.7, and API and TSI, for three
cities as example. The number of correlation is N=4607 points.
Thus the PCC $r\geq0.038$ could be considered as significant under
the significance level of $\alpha=0.01$ (Confidence level is
$99\%$). For Beijing city, the PCC are small but higher than
0.038. The SCC and KCC are also small but significant with
two-sided significance of its deviation near zero. This indicate
the weak but believable correlation between API and SSN, API and
F10.7, API and TSI. Beijing city has more than $40.4\%$ windy days
(wind scale is $>3$, wind speed is $V>3.4m/s$). It also suffer the
sand storm or dust storm sometimes. Thus air particles of Beijing
city might have larger percent of dust than other cities with few
sand storm or dust storm. These might influence the API
moderately. For Chengdu city, the absolute value of PCC are lower
than 0.038, and also lower than SCC and KCC. The API of Chengdu
city is influenced slightly by the wind and dust with only $7.7\%$
windy days ($V>3.4m/s$) and no dust storm. For Kunming city, all
correlation coefficient are negative. The absolute value of PCC
are higher than 0.038, and also higher than absolute value of SCC
and KCC. The API of Kunming city is also influenced slightly by
the wind and dust with only $7.7\%$ windy days ($V>3.4m/s$) and no
dust storm. The correlation between API and SSN, API and F10.7,
and API and TSI is weak and quite complex in a single city. It
might be positive, negative, and varied near zero. We will analyze
them systematically in subsection 2.4.

Fig.\ref{figsca} plotted for three cities the scatter diagram of
API versus SSN, API versus F10.7, API versus TSI, and percent of
large API, respectively. Upper part of panel (a) plotted the
scatter diagram of API versus SSN. The red cross is the API
average value at given SSN ($\overline{API_{SSN}}$). The
$\overline{API_{SSN}}$ is varied and increased along with
$SSN\leq180$, then reach the maximum and is decreased along with
$SSN\geq180$. The $\overline{API_{SSN}}$ around SSN=180 is about
two times as large as the $\overline{API_{SSN}}$ of $SSN\leq50$.
Bottom panel of (a) plotted the percent of large API
($>\overline{API}+2\sigma$). It is increased along with
$SSN\leq150$, then reach the maximum around $150\leq SSN\leq180$,
and no large API when $SSN\geq180$. The highest percent of large
API around $SSN=150$ is about four times as large as that of
$SSN\leq90$. The results of $SSN>200$ should be taken care with
only $0.56\%$ points. Panel (b) and (c) showed the similar result
of API versus F10.7 and API versus TSI. Panel (d), (e), and (f)
showed the results of Chengdu city. The $\overline{API_{SSN}}$ and
percent of large API have no significant difference along with the
SSN, F10.7, and TSI. Panel (g), (h), and (i) showed the similar
results of Kunming city.

The AQI of the single city have also been studied. But the result
is ambiguous and uncertain since only several hundreds of days of
observation. The global result of AQI will be showed in next
subsection.

\subsection{The global probability distribution of API and AQI in Chinese big cities}

\begin{figure*}[ht]
\epsscale{1.50} \plotone{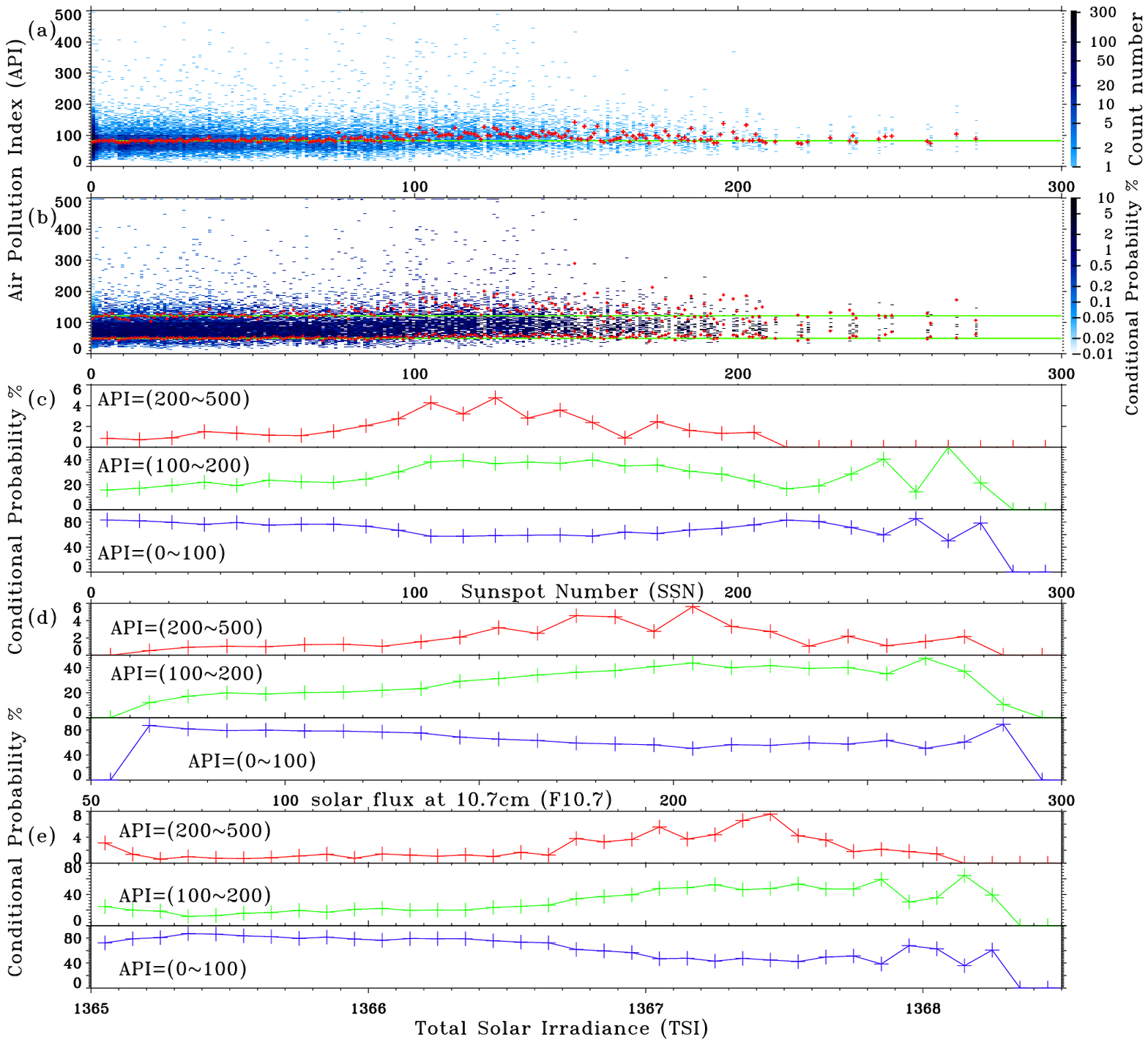} \caption{Global probability
distribution of API of the 14 candidate cities with $\geq0.55\%$
days of large API ($\geq200$). Panel (a) is the count number of
API along SSN. The red circle is the API average at given SSN. The
green line is the API average of $SSN\leq50$; Panel (b) is
$P(API|SSN)$ with bin size of 1 SSN. The red circle is the
boundary of $10\%$ large API and $10\%$ low API. The two green
lines are the $10\%$ boundary given $SSN<50$; Panel (c) plotted
$P(API|SSN)$ of three division of API given that SSN with bin size
of 10; Panel (d) plotted $P(API|F10.7)$ of three division of API
given that F10.7 with bin size of 10; Panel (e) plotted
$P(API|TSI)$ of three division of API given that TSI with bin size
of 0.1. \label{highapi}}
\end{figure*}


There are 120 Chinese cities had measured the API during
2000-2013. The city with insufficient data will be excluded since
it will blur the results. The data of two cities of Lanchow and
Urumchi were also excluded because the air particle of these two
cities have larger percent of sand storm or dust storm than other
cities. The sand storm or dust storm will influence the API too
much. The solar activity can take almost nothing effect on sand
storm or dust storm because it appeared suddenly and intensively.
Thus, we only analyze the data of 45 cities with successive data
of more than 4000 (Max=4938) days and with few sand storm or dust
storm. Among them 14 cities have more than $>0.55\%$ days with
large API ($\geq200$); 16 cities have $0.1\%\sim0.55\%$ days with
large API ($\geq200$); 15 cities have less than $<0.1\%$ days with
large API ($\geq200$). We will study the global probability
distribution and conditional probability of API of 3 classes of
Chinese big cities. The conditional probability $P(API|SSN)$,
$P(API|F10.7)$, and $P(API|TSI)$ were mainly analyzed. The
$P(AQI|SSN)$, $P(AQI|F10.7)$ and $P(AQI|TSI)$ were also analyzed.
Globally, during 2000-2013, the probability $P(SSN<100)$,
$P(100\leq SSN<200)$, and $P(SSN\geq200)$ are about $84.3\%$,
$15.1\%$, and $0.56\%$, respectively. For all the 45 candidate
cities, the probability $P(API<100)$, $P(100\leq API<200)$, and
$P(API\geq200)$ are about $87\%$, $13.1\%$, and $0.87\%$,
respectively.

\begin{figure*}[ht]
\epsscale{1.50} \plotone{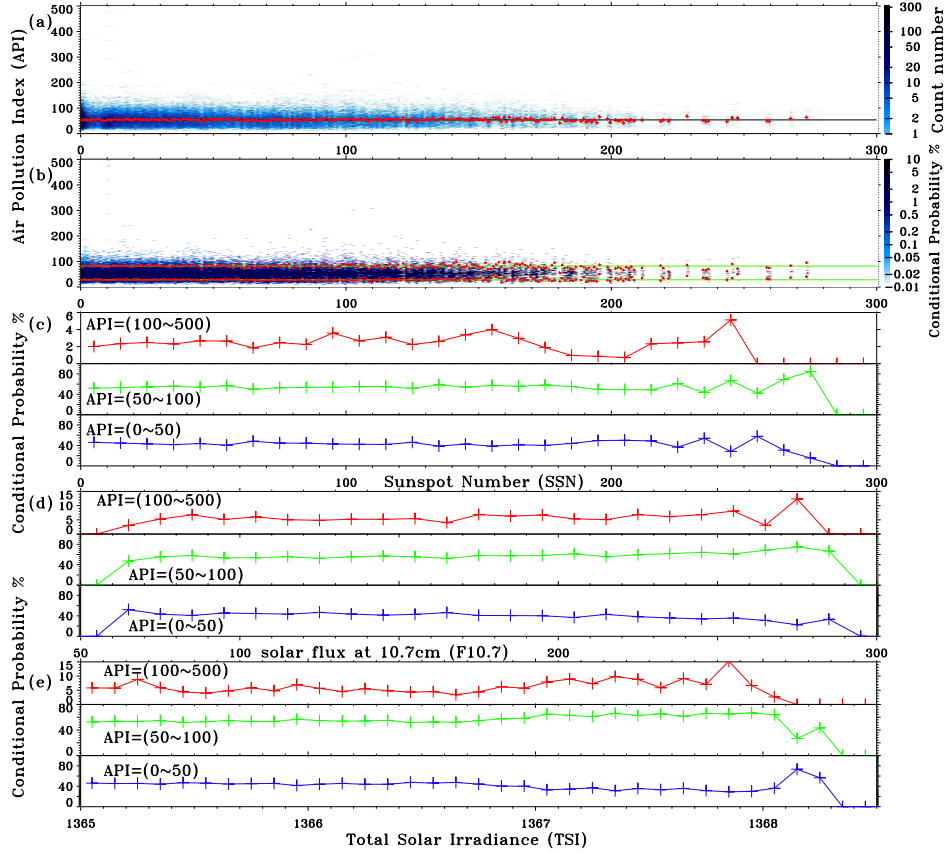} \caption{Global probability
distribution of API of the 15 candidate cities with $<0.1\%$ days
of large API ($\geq200$). Panel (a) is the count number of API
along SSN. The red circle is the API average at given SSN. The
green line is the API average of $SSN\leq50$; Panel (b) is
$P(API|SSN)$ with bin size of 1 SSN. The red circle is the
boundary of $10\%$ large API and $10\%$ low API. The two green
lines are the $10\%$ boundary given $SSN<50$; Panel (c) plotted
$P(API|SSN)$ of three division of API given that SSN with bin size
of 10; Panel (d) plotted $P(API|F10.7)$ of three division of API
given that F10.7 with bin size of 10; Panel (e) plotted
$P(API|TSI)$ of three division of API given that TSI with bin size
of 0.1. \label{lowapi}}
\end{figure*}

Fig.\ref{highapi} plotted the API probability distribution of the
14 cities with more than $0.55\%$ days of large API ($\geq200$).
Panel (a) plotted the count number distribution of API along SSN.
Globally, the probability $P(API<100)$, $P(100\leq API<200)$, and
$P(API\geq200)$ are about $77.4\%$, $22.2\%$, and $1.5\%$,
respectively. Given that $SSN<100$, the probability $P(API<100)$,
$P(100\leq API<200)$, and $P(API\geq200)$ are about $77.0\%$,
$21.6\%$, and $1.4\%$, respectively. Given that $100\leq SSN<200$,
the probability $P(API<100)$, $P(100\leq API<200)$, and
$P(API\geq200)$ are about $61.3\%$, $36.0\%$, and $2.7\%$,
respectively. Given that $200\leq SSN<300$, the probability
$P(API<100)$, $P(100\leq API<200)$, and $P(API\geq200)$ are about
$73.2\%$, $26.7\%$, and $0.002\%$, respectively. There is no
$API\geq200$ given that $SSN\geq210$. But it is suspicious to
concluded that $P(API\geq200|SSN\geq210)$ will be zero with only a
very small proportion $0.39\%$ of the sample of one solar cycle.
The red circle is the API average at given SSN. It is varied and
increased along with $SSN\leq150$, then reach the maximum and then
is decreased along with $SSN\geq195$. The API average around
SSN=150 is about 1.72 times of the API average of $SSN\leq50$
which showed as green line in panel(a). All these indicated that
the API will be more likely higher around certain value of SSN
than others. In order to ascertain this, the conditional
probability $P(API|SSN)$ should be analyzed in detail. Panel (b)
plotted $P(API|SSN)$ given that SSN with bin size of 1. The
$P(API|SSN)$ around ($100\leq SSN<150$) is a little more likely to
be large $API\geq200$ than others. The red cross in panel (b) is
the boundary of $10\%$ large API and $10\%$ low API. The two green
lines are the average value of $10\%$ boundary of large API and
low API given $SSN<50$, respectively. The boundary of $10\%$ large
API and $10\%$ low API varied similar as API average in panel (a).

Panel (c) plotted $P(API|SSN)$ given that SSN with bin size of 10.
For each bin of given SSN, the API was divided into three
divisions: $API<100$, $100\leq API<200)$, and $API\geq200$. When
$SSN\geq210$, the $P(API|SSN)$ of three division varied too much
because only $\simeq0.39\%$ data are in the range of $SSN\geq210$.
When $SSN\leq210$, the $P(API_{100\sim200}|SSN)$ and
$P(API\geq200|SSN)$ around $100\leq SSN\leq150$ are higher than
that of other given SSN; while the $P(API<100|SSN)$ around
$100\leq SSN\leq150$) are lower than that of other given SSN. The
maximum $P(API_{200\sim500}|SSN)$ given $120\leq SSN<130$ is about
5.7 times of $P(API_{200\sim500}|SSN)$ given low SSN. In panel (d)
of fig.\ref{highapi}, the $P(API\geq 200/F10.7)$ around $200\leq
F10.7\leq210$ are higher than that of others given F10.7. The
maximum $P(API\geq200|F10.7)$ given $200\leq F10.7<210$ is about
5.2 times of $P(API\geq200|F10.7)$ given low F10.7. In panel (e)
of fig.\ref{highapi}, the $P(API\geq200|TSI)$ around $1367.4\leq
TSI\leq1367.5$ are higher than that of others given TSI. The
maximum $P(API\geq200|TSI)$ given $1367.4\leq TSI<1367.5$ is about
7.2 times of $P(API\geq200|TSI)$ given $1365.2<TSI<1366.5$. The
relationship between API and SSN, API and F10.7, and API and TSI
are weak but clear. The probability of high API will be increased
first, then reach to a maximum, and then decreased along with the
increasing of SSN, F10.7 or TSI. We did not plotted the API
probability distribution of the 16 cities with $0.1\%\sim0.55\%$
days of large API ($\geq200$). The result is similar as
fig.\ref{highapi}. But when $API=(100\sim500)$, the $P(API|SSN)$
given $100\leq SSN\leq150$ are smaller than that of
fig.\ref{highapi}.

\begin{figure*}[ht]
\epsscale{1.50} \plotone{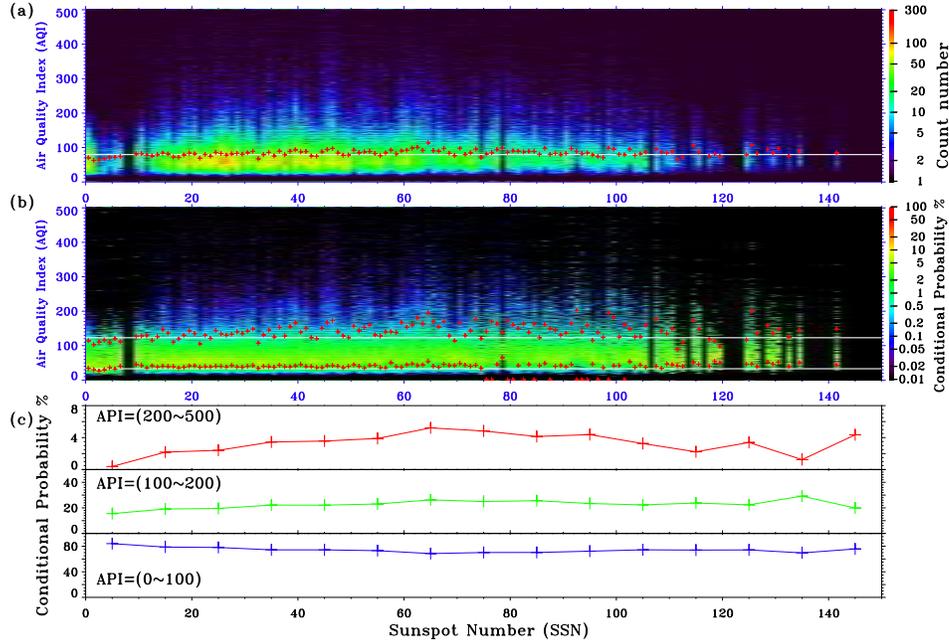} \caption{Global probability
distribution of AQI of most Chinese big cities. Panel (a) is the
count number of AQI along SSN; Panel (b) is $P(AQI|SSN)$ with bin
size of 1 SSN; Panel (c) plotted $P(AQI|SSN)$ of three division of
AQI given that SSN with bin size of 10. \label{figaqi}}
\end{figure*}

Fig.\ref{lowapi} plotted the API probability distribution of the
15 cities with less than $0.1\%$ days of large API ($\geq200$).
Panel (a) plotted the count number distribution of API along SSN.
Globally, the probability $P(API<100)$, $P(100\leq API<200)$, and
$P(API\geq200)$ are about $97.6\%$, $2.3\%$, and $0.02\%$,
respectively. The API average (red cross) are about the same for
all cases of SSN. The Panel (b) plotted $P(API|SSN)$ given that
SSN with bin size of 1. The boundary of $10\%$ large API and
$10\%$ low API are also about the same for all cases of SSN. Panel
(c) plotted $P(API|SSN)$ given that SSN with bin size of 10. For
each bin of given SSN, the API was divided into three divisions:
$API<100$, $100\leq API<200)$, and $API\geq200$. When
$SSN\geq210$, the $P(API|SSN)$ of three division also varied too
much because only $\simeq0.39\%$ data. When $SSN\leq210$, the
$P(API_{100\sim500}|SSN)$ around $90\leq SSN\leq160$) are a little
higher than that of other given SSN; while $P(API<100|SSN)$ are
about the same for all cases of given SSN. Panel (d) and (f)
showed that $P(API|F10.7)$ and $P(API|TSI)$ are about the same for
all cases of given $F10.7\leq260$ and $TSI\leq1367.8$. Comparing
with fig.\ref{highapi}, the SSN (or F10.7, or TSI) take weaker
impact on the cities with low API than the cities with high API.

There are 367 Chinese cities had measured the AQI during
2014-2016. The short period of measurement are not sufficient for
data analyze of relationship. Only small amount of cities have a
small quantity of measured data. Most of the cities with
successive data of more than 600 (Max=942) days will be analyzed.
Fig.\ref{figaqi} plotted the AQI probability distribution of these
cities. Globally, the probability $P(AQI<100)$, $P(100\leq
AQI<200)$, and $P(AQI\geq200)$ are about $74.1\%$, $22.5\%$, and
$3.4\%$, respectively. The $P(AQI\geq200)$ about 2.2 times as high
as $P(API\geq200)$ of 15 large API cities. AQI add the index of
suspended particulates smaller than 2.5 $\mu$m in aerodynamic
diameter (PM2.5). While API of most Chinese cities did not add the
index of PM2.5. Thus AQI is lager than API in a city. Panel (a)
plotted the count number distribution of AQI along SSN. The AQI
average (red cross) at given SSN is also varied and increased
along with $SSN\leq70$, then reach to a maximum and then go on
varying along with $SSN\geq70$. The API average around SSN=150 is
about 1.72 times of the API average of $SSN\leq50$ which showed as
green line in panel(a). The Panel (b) plotted $P(AQI|SSN)$ given
that SSN with bin size of 1. The boundary of $10\%$ large AQI and
$10\%$ low AQI (red cross) is also varied and increased along with
$SSN\leq70$, then reach to a maximum and then go on varying along
with $SSN\geq70$. Panel (c) showed that the $P(AQI|SSN)$ of large
$AQI\geq200$ reach the maximum at given $SSN=60\sim70$. But the
$P(AQI|SSN)$ of large $AQI\geq200$ are also high at given
$SSN=140\sim150$. This is a little difference comparing
fig.\ref{highapi} and fig.\ref{lowapi}. Please note that only
$<5\%$ cases are in the range of $SSN\geq100$ during 2014-2016.

\begin{figure*}[ht]
\epsscale{1.50} \plotone{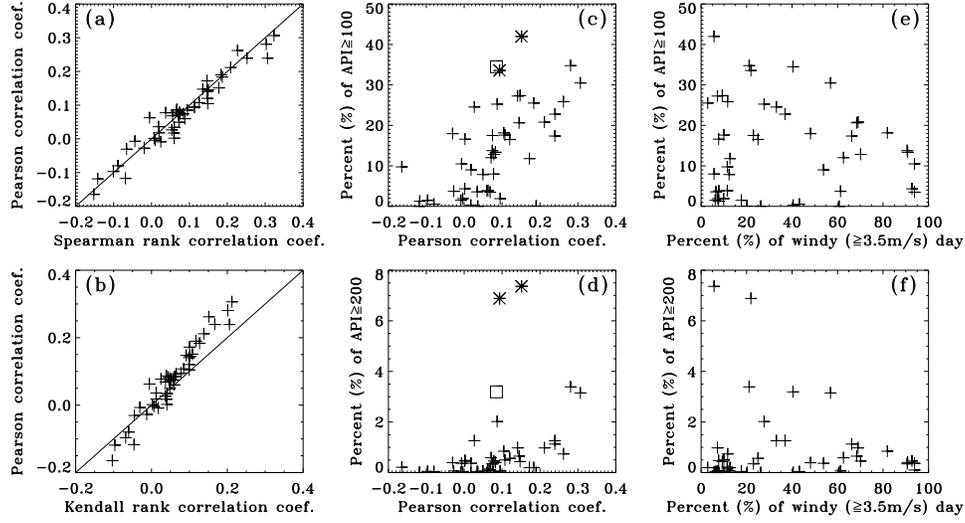} \caption{The relationship
among different correlation, API level, and wind. Panel (a) and
(b) are Pearson correlation coefficient versus Spearman rank
correlation coefficient and Pearson correlation coefficient versus
Kendall rank correlation coefficient, respectively. Panel (c) and
(d) plotted the percent of API level versus the Pearson
correlation coefficient. The star is the value of Lanchow and
Urumchi cities. The square is the value of Beijing city. Panel (e)
and (f) plotted the percent of API level versus the windy days.
\label{apicor}}
\end{figure*}

\subsection{The relationship among different correlation,
API level, and wind for most Chinese big cities}

For all the 45 candidate cities, we studied the Pearson
correlation, Spearman rank correlation, and Kendalls rank
correlation of API and SSN. Panel (a) of fig.\ref{apicor} shows
the scatter diagram of PCC versus SCC. The difference between PCC
and SCC is small and varied symmetrically within a small linear
range. Panel (b) of fig.\ref{apicor} shows that the absolute value
of PCC is higher than that of KCC globally. Thus we can conclude
that the correlation between API and SSN is linearly, that is
Pearson correlation. We also studied the PCC, SCC, and KCC of API
versus F10.7 and API versus TSI. The correlation of API and F10.7,
API and TSI are also linearly (Pearson correlation).

The relationship between API level and correlation are studied.
Panel (c) of fig.\ref{apicor} shows the scatter diagram between
percent of $API\geq100$ and PCC of API/SSN. The higher percent of
large API $\geq100$, the higher correlation. Panel (d) of
fig.\ref{apicor} also shows higher percent of large API $\geq200$,
the higher correlation coefficient, except that two cities of
Lanchow and Urumchi (star symbol). As discussed in subsection 2.3.
The air particle of Lanchow and Urumchi are influenced by sand
storm or dust storm which appeared suddenly and intensively. The
relationship between API level and wind are plotted at panel (e)
and (f) of fig.\ref{apicor}. The percent of large API will be
smaller if the percent of windy days ($\geq3.4m/s$) is larger. It
is easy to understand that the wind blow away the pollution
particles.

\subsection{The total solar irradiance and sunspot number}

\begin{figure*}[ht]
\epsscale{1.50} \plotone{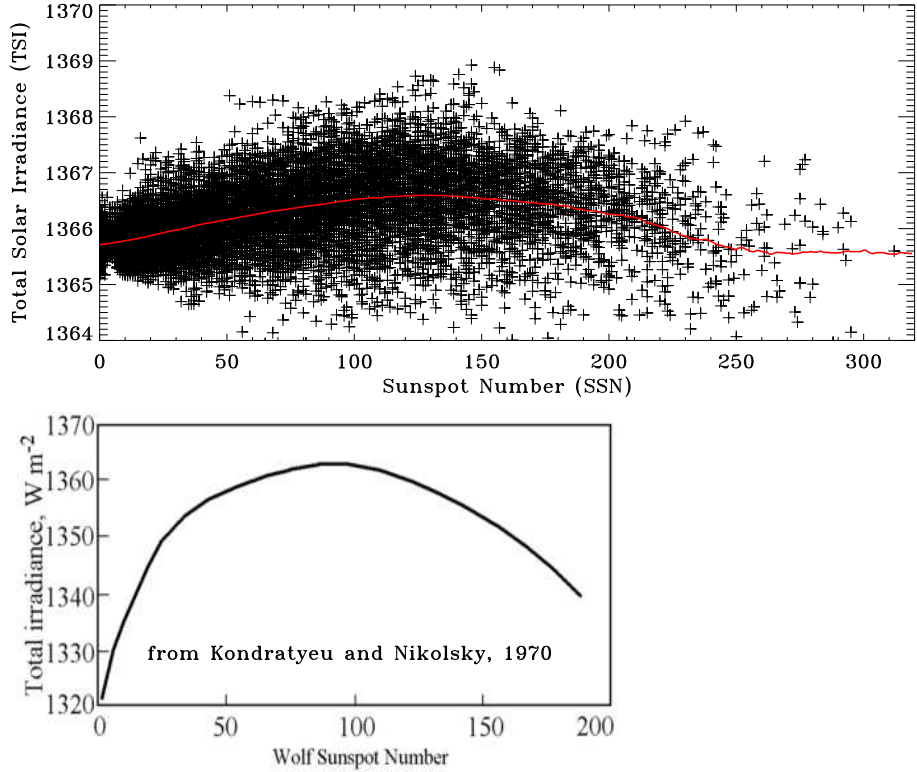} \caption{Upper panel is the
scatter diagram of TSI versus SSN. Bottom panel is the result of
\citep{Kondrat70}.\label{tsissn}}
\end{figure*}

The relationship between TSI and SSN is quite certain. The cross
correlation between TSI and SSN shows that maximum correlation
happened when SSN is 29 days before TSI. The upper panel of
fig.\ref{tsissn} shows the scatter diagram of TSI versus SSN. The
red line is the moving average of the mid value of TSI versus SSN.
It shows that the TSI increased globally when $SSN<100$, then
reach to a maximum around $100\leq SSN\leq150$, and then is
decreased when $SSN>150$. This is a little difference than the
result of \citep{Kondrat70} as showed in bottom panel of
fig.\ref{tsissn}. However, the SSN is not higher than 200 in the
result of \citep{Kondrat70}.

\section{Conclusion and Discussion}

The air pollution should be produced mainly from industry,
automobiles and domestic fuels fire, cooking, smoking and so on.
There is no doubt that weather condition take the major effect on
the API (or AQI): Fog, wind, cloud, raining, humidity,
temperature, etc. Solar activity can affect the solar irradiance
and ionization in the atmosphere, thus impact on the ozone
molecule, the lower atmosphere state, weather condition, and urban
environment on the earth. With the data analysis between API (or
AQI) of hundreds of Chinese cities and SSN (or F10.7, or TSI), we
study the relationship among them and find some preliminary
results as follows.

For each city, the correlation coefficient (PCC, NCC, KCC) between
API and SSN (or F10.7, or TSI) are weak $-0.17<r<0.32$ and
complex, considering a sample of one solar cycle ($>4000$ points).
It might be positive, negative, and varied near zero. For the
Beijing city with high air pollution, both the API average at
given SSN and percent of large API at given SSN are varied and
increased along with SSN, then reach to a maximum around $150\leq
SSN\leq180$ and then are decreased. The API average around SSN=180
is about two times as large as the API average of $SSN\leq50$, and
the percent of large API around $SSN=150$ is about four times as
large as that of $SSN\leq90$. For the city with moderate or low
air pollution, The API average and percent of large API have no
significant difference along with the SSN. The relationship
between API and F10.7, and API and TSI are similar.

Globally, during 2000-2013, the probability $P(SSN<100)$,
$P(100\leq SSN<200)$, and $P(SSN\geq200)$ are about $84.3\%$,
$15.1\%$, and $0.56\%$, respectively. For all the 45 candidate
cities, the probability $P(API<100)$, $P(100\leq SSN<200)$, and
$P(API\geq200)$ are about $87\%$, $13.1\%$, and $0.87\%$,
respectively. For the 14 high polluted cities: the
$P(API\geq200|SSN)$ around $100\leq SSN\leq150$ are higher than
that of other given SSN, and the maximum $P(API\geq200|SSN)$ given
$120\leq SSN<130$ is about 5.7 times of $P(API\geq200|SSN)$ given
low SSN; the $P(API\geq200|F10.7)$ around $200\leq F10.7\leq210$
are higher than that of others given F10.7, and the maximum
$P(API\geq200|F10.7)$ given $200\leq F10.7<210$ is about 5.2 times
of $P(API\geq200|F10.7)$ given low F10.7; the $P(API\geq200|TSI)$
around $1367.4\leq TSI\leq1367.5$ are higher than that of others
given TSI, and the maximum $P(API\geq200|TSI)$ given $1367.4\leq
TSI<1367.5$ is about 7.2 times of $P(API\geq200|TSI)$ given low
$TSI=1365.2\sim1366.5$. The relationship between API and SSN, or
API and F10.7 or API and TSI are weak but clear. The probability
of high API will be increased first, then to a maximum, and then
decreased along with SSN, F10.7 or TSI. The result of middle
polluted cities are similar but the $P(API\geq200|SSN)$ around
$100\leq SSN\leq150$ are smaller than that of high polluted
cities. The SSN (or F10.7, or TSI) take weaker impact on the
cities with lower API than the cities with higher API.

The relationship between AQI and SSN, AQI and F10.7, and AQI and
TSI are also weak but a little difference than the relationship
between API and SSN, API and F10.7, and API and TSI. It is
difficult to give more conclusion now since only a small sample of
data during 2014-2016. And the AQI have only $<5\%$ cases in the
range of $SSN\geq100$ and no cases in the range of $SSN\geq150$.

Comparing the Pearson correlation, Spearman rank correlation, and
Kendalls rank correlation, Pearson correlation coefficient is the
largest globally. That is, the correlation of API and SSN is
linearly. The relationship between API level and PCC indicated
that the higher percent of large API, the higher correlation
coefficient. The relationship between API level and wind indicated
that the percent of large API will be smaller if the percent of
windy days ($\geq3.4m/s$) is larger.

The question brought out in the introduction can be answer now.
The solar activity take direct effect on the API and AQI on the
earth through solar radiation \citep{Muda12} and high energy
particles. The solar activity also take indirect and long term
effect on lower atmosphere and weather, thus the API and AQI on
the earth. The relationship between API and solar activity is weak
and complex. The solar activity does not impact on air pollution
monotonously. The probability of high API will be increased first,
then to a maximum, and then decreased along with the solar
activity. Former works \citep{PAP85} on solar irradiance and
sunspot area shows that strong inverse correlation is shown
between the irradiance observed by the ACRIM radiometer and the
projected areas of the 'active' sunspots. \citep{Hemp12} concluded
that the Earth atmosphere acts as an amplifier between space and
ground, and that the amplification is probably controlled by solar
activity. They suspected the cosmic rays intensity as the link
between solar activity and atmospheric transparency. Our
preliminary study did not consider the influence of the charged
particles on the earth surface. The larger solar activity will
result more charged particles thus congregate the polluted
particles, thus reduce the number of polluted particles.

The solar activity will influence the air pollution. This
influence is weak and complex because man-made environment and
weather condition take the dominant influence. We can anticipate
that the result will be more clear with a sample of more than one
solar cycle.

 \acknowledgments The author thanks the referee for helpful and
valuable comments on this paper. This work is supported by NSFC
grants 11373039, 11433006, 11573039, and 11661161015, the MOST
grant 2014FY120300, the Specialized Research Fund for State Key
Laboratories of Space Weather. Thanks to NGDC, ACRIMSAT/ACRIM,
SORCE/TIM, and China's MEP for the online data.

\section{Citing references}

\smallskip
\noindent
\verb!\citep{Cooper12}! -- \citep{Cooper12}\\
\verb!\citep{Hemp12}! -- \citep{Hemp12}\\
\verb!\citep{Kondrat70}! -- \citep{Kondrat70}\\
\verb!\citep{Kopp05a}! -- \citep{Kopp05a}\\
\verb!\citep{Kopp05b}! -- \citep{Kopp05b}\\
\verb!\citep{Kutiev13}! -- \citep{Kutiev13}\\
\verb!\citep{Mavrakis08}! -- \citep{Mavrakis08}\\
\verb!\citep{Muda12}! -- \citep{Muda12}\\
\verb!\citep{PAP85}! -- \citep{PAP85}\\
\verb!\citep{Pudovkin04}! -- \citep{Pudovkin04}\\
\verb!\citep{Sharma97}! -- \citep{Sharma97}\\
\verb!\citep{Sicard11}! -- \citep{Sicard11}\\
\verb!\citep{Willson14}! -- \citep{Willson14}\\


\end{document}